\documentclass[aps,prb,floatfix,twocolumn]{revtex4}
\usepackage{graphicx}
\usepackage{amsmath}
\usepackage{amssymb}
\newcommand{\shalf}{{{\tiny{\frac{1}{2}}}}}
\newcommand{\BEQ}{\begin{equation}}
\newcommand{\EEQ}{\end{equation}}
\newcommand{\BEA}{\begin{eqnarray}}
\newcommand{\EEA}{\end{eqnarray}}

\newcommand{\as}{{\rm assistant}}

\newcommand{\al}{\alpha}

\newcommand{\tr}{{\rm tr}}

\newcommand{\sz}{\hat{s}_3}
\newcommand{\sy}{\hat{s}_2}

\newcommand{\sx}{\hat{s}_1}
\newcommand{\siz}{\hat{\sigma}_3}
\newcommand{\siy}{\hat{\sigma}_2}

\newcommand{\six}{\hat{\sigma}_1}

\newcommand{\hH}{\hat{H}} 
\newcommand{\hrho}{\hat{\rho}} 
\newcommand{\hR}{\hat{{\cal R}}} 
 
\newcommand{\hO}{\hat{\Omega}} 
\newcommand{\ho}{\hat{\omega}} 
\newcommand{\hpi}{\hat{\pi}} 
\newcommand{\hU}{\hat{U}}

\newcommand{\hr}{\hat{r}} 

\newcommand{\half}{\frac{1}{2}}
\begin{document} 

\title{Determining a quantum state by means of a single apparatus.}

\author{A.E. Allahverdyan$^{1,2)}$, R. Balian$^{3)}$
and Th.M. Nieuwenhuizen$^{1)}$}
\affiliation{$^{1)}$ Institute for Theoretical Physics,
Valckenierstraat 65, 1018 XE Amsterdam, The Netherlands}
\affiliation{$^{2)}$Yerevan Physics Institute,
Alikhanian Brothers St. 2, Yerevan 375036, Armenia}
\affiliation{$^{3)}$ SPhT, CEA-Saclay, 91191 Gif-sur-Yvette cedex, France}

\begin{abstract}
The unknown state $\hrho$ of a quantum system S is determined by
letting it interact with an auxiliary system A, the initial state of
which is known.  A one-to-one mapping can thus be realized between the
density matrix $\hrho$ and the probabilities of occurrence of the
eigenvalues of a single and factorized observable of S+A, so that
$\hrho$ can be determined by repeated measurements using a single
apparatus. If S and A are spins, it suffices to measure simultaneously
their $z$-components after a controlled interaction. 
The most robust setups are determined in this case, 
for an initially pure or a completely disordered state of A. 
They involve an Ising or anisotropic Heisenberg coupling and 
an external field.

\end{abstract}



\maketitle

Consider a set of identical quantum systems S, prepared in some
unknown state $\hrho$. Information on $\hrho$ can be gathered by
measuring some observable $\ho$ of S. Such a measurement has a
statistical nature, whether $\hrho$ is pure or not: when repeated many
times, it provides the probabilities $p_i=\tr\hrho\hpi_i$ for the distinct
eigenvalues $\omega_i$ of an operator $\ho$, where $\hpi_i$ are the associated
eigenprojections. We thus find partial information on $\hrho$. A question then
arises which lies at the heart of quantum theory: Which observables
need to be measured for a complete determination of $\hrho$
~~\cite{kem,hist,leon,wei}?  The recognition that certain {\it
non-commuting} observables have to be measured for that purpose was
used by Bohr to formulate the principle of complementarity
\cite{kem,bohr}.  Later on the problem of determining an unknown state
was considered from various perspectives for continuous \cite{leon}
and discrete systems \cite{wei} and found applications in quantum
communication \cite{helstrom,reviews}. However, during all these
developments it was not questioned whether non-commutative
measurements are truly needed.

Here we face the problem of determining from some collection of
experimental data the whole set of unknown matrix elements of the
state $\hrho$ of a system S. In fact, what we call ``the state of a
system'' refers to a density matrix describing an {\it ensemble}, 
and the determination of the
data will require repeated experiments. However, we will show that it
is sufficient for our purposes to produce these data by means of a
{\it single apparatus} which measures commuting observables only.  The
key of the method consists in coupling the system S to an auxiliary
system A, the state of which we know. We will show that the full
$\hrho$ can then be deduced from simultaneous measurements of two
obviously commuting observables, $\hat{\omega}$ pertaining to S and
$\hat{o}$ pertaining to A, respectively.  Repeated measurements of
$\hat{\omega}$ and $\hat{o}$ 
yield the joint probabilities of occurrence for all pairs of eigenvalues
of  $\hat{\omega}$ and $\hat{o}$; this
will be sufficient to determine the whole density matrix $\hrho$ of S,
provided the initial state of A and the joint evolution of S+A are
known.  We shall discuss the most robust measurements of this type and
point out certain advantages that this scheme provides over the usual
methods of state determination. Realistic experiments can be designed
along these lines, as we shall see.

{\it General reasoning.} The following counting argument already suggests 
the feasibility of the above idea: The number of real parameters to be
determined for finding $\hrho$ which lives in $m$-dimensional Hilbert
space is $m^2-1$, since $\hrho$ is a $m\times m$ hermitian matrix with
unit trace. On the other hand, the most informative repeated
measurements of an observable $\ho$ of S are those for which the
spectrum $\omega_i$ is non-degenerate; they provide $m-1$ independent
data, the probabilities $\tr \hrho\hpi_i$, $i\leq 1\leq m$. Thus, if
measurements are performed on S only, they must deal with at least
$(m^2-1)/(m-1)=m+1$ non-commuting observables in order to fully determine the
unknown $\hrho$.  For instance, the $2\times 2$ density matrix
$\hrho=\shalf(\hat{1}+\vec{\rho}\cdot\vec{\hat{\sigma}})$
of a spin-$\shalf$ is parameterized by the $m^2-1=3$ expectation values
$\vec{\rho}={\rm tr} (\,\hrho\cdot\vec{\hat{\sigma}}\,)$ of the Pauli
matrices $\vec{\hat{\sigma}}=(\hat{\sigma}_1,\hat{\sigma}_2,
\hat{\sigma}_3)$ with $\hat\sigma_1\hat\sigma_2=i\hat\sigma_3$,
 and its determination requires measuring the spin in
$m+1=3$ non-coplanar directions. 
Experiments with three different apparatuses are thus needed. 
This conclusion holds even if we are given the information 
that the unknown state $\hrho$ is pure 
($\hat\rho\,^2=\hat\rho$ or $\vec{\rho}\,^2=1$), 
since then the measurement of two components of $\vec\rho$
does not fix the sign of the third one.
For $m\geq 3$ the construction of the needed $m+1$ non-commuting
observables is less straightforward, see
the discussion in Ref.~\cite{wei}.

In order to design a scheme where $\hrho$ will be determined by
measurements of a single observable (or, equivalently, by commuting
measurements only), it is natural to introduce ~\cite{Ariano} 
an auxiliary system A that we term {\it the assistant}, and which lies 
in a {\it known state} $\hr$.  Let
$n$ be the number of dimensions of the Hilbert space of A.  The
compound system S+A has a $mn\times mn$ density matrix
$\hR=\hrho\otimes\hr$. 
Ref. ~\cite{Ariano}  proposes measuring {\it one} of its
observables, a ``universal quantum observable'',
$\hO=\sum \Omega_\alpha\hat{P}_\alpha$, where the
spectrum $\Omega_\alpha$ is not degenerate so that the
eigenprojections $\hat{P}_\alpha$ (with $1\leq \alpha\leq mn$, noting that
$\sum_\alpha\hat{P}_\alpha=\hat{1}$) constitute a complete set of
$mn-1$ commuting observables. Such repeated measurements provide
$mn-1$ independent data $P_\alpha= \tr\hR\hat{P}_\alpha$,
the probabilities of the eigenvalues $\Omega_\alpha$ of $\hO$. The set
$P_\alpha$ also represents the diagonal elements of the density matrix
$\hrho\otimes\hr$ in the basis that diagonalizes $\hO$. A {\it linear
mapping} $\hrho\mapsto P_\alpha= \sum_{ij}R_{\al,ij}\rho_{ji}$ is thus
generated, leading from the $m^2-1$ real parameters of $\hrho$ to the
$mn-1$ data $P_\alpha$. If the \as ~A has the same dimensionality $m$
as S, this mapping is represented by a square matrix. In general, if
the measured observable $\hO$ intertwines sufficiently S and A, the
determinant $\Delta={\rm det}\,R_{\alpha,ij}$ of this matrix may be
expected to be non-zero.  The inverse mapping then
solves our problem: measurements of $\hO$, performed repeatedly with a
single apparatus, yield the probabilities $P_\alpha$, the knowledge of
which is equivalent to that of $\hrho$. If $n>m$, the set $P_\alpha$
can still determine $\hrho$, but it is overabundant.

However, the above scheme is not easy to implement in practice, since it
implies measuring an observable $\hO$ (or a commuting set
$\hat{P}_\alpha$) which thoroughly mixes S and A. We propose here a
modified procedure, which will allow a much simpler choice for $\hO$.
The measurement of $\hO$ is performed not at the time $t=0$, at 
which S is prepared in the unknown state $\hrho$ and A in the known 
state $\hr$, but at a later time $t=\tau$.  During the
lapse $0<t<\tau$, S and A interact, their evolution being generated by
a {\it known} Hamiltonian $\hH$. The state of the composite system S+A
which is tested is now $\hR_\tau=\hU \hR_0\hU^\dagger$, where the
initial state is $\hR_0=\hrho\otimes\hr$ and the evolution operator is
$\hU={\rm e}^{-{\rm i}\hH\tau}$. The required {\it mixing} of $\hrho$ and
$\hr$ being thus {\it achieved by dynamics}, we can now measure the simplest
possible non-degenerate observable $\hO$, a factorized
quantity $\hO=\hat{\omega}\otimes\hat{o}$. The observables
$\hat{\omega}$ and $\hat{o}$ of S and A have the spectral
decompositions $\hat{\omega}=\sum_{i=1}^m\omega_i\hpi_i$ and
$\hat{o}=\sum_{a=1}^n o_a\hat{p}_a$ and the projection operator
$\hat{P}_\alpha$, with $\alpha=\{ia\}$, takes the form
$\hat{P}_\alpha\equiv\hat{P}_{ia}=\hpi_i \otimes \hat{p}_a$. Repeated
measurements of $\hO$, that is, of $\hat{\omega}$ and $\hat{o}$
simultaneously, determine the joint probabilities
\BEA
P_\alpha\equiv P_{ia}=\tr \,\hU(\hrho\otimes\hr)\hU^\dagger(\hat{\pi}_i
\otimes\hat{p}_a)
\label{2}
\EEA 
to observe $\omega_i$ for S and $o_a$ for A. (The numbers $P_{ia}$
are the diagonal elements of $\hU(\hrho\otimes\hr)\hU^\dagger$ in the
factorized basis which diagonalizes $\hat{\omega}$ and $\hat{o}$.) 
Like above, the mapping $\hrho\mapsto P_{\alpha}$ is expected to be invertible 
for $n\geq m$, provided $\hH$ couples S and A sufficiently.  Then {\it
simultaneous measurements} of $\hat{\omega}$ on the system S and of
$\hat{o}$ on the \as ~A, based on the counting of the events
$\{ia\}$ and of their correlations, fully determine $\hrho$ through 
inversion of the equation (\ref{2}).

For given observables $\hat{\omega}$ and $\hat{o}$ and for a given
initial state $\hr$ of the \as, the precision of this scheme of
measurement of $\hrho$ relies on the ratio between the experimental
uncertainty about the set $P_{\alpha}$ and the resulting uncertainty on
$\hrho$, which can be characterized by the determinant $\Delta$ 
of the transformation (\ref{2}). For $\Delta=0$ it would be
impossible to determine $\hat{\rho}$ by means of $P_{\alpha}$.  The
Hamiltonian $\hH$ and the duration $\tau$ of the interaction should
thus be chosen so as to maximize $|\Delta|$.

{\it Two by two density matrix.} In the following we illustrate the
above ideas by studying a two-level system S ($m=2$). We exhibit in
particular the best measurement schemes that correspond to the largest
$|\Delta|$. We use the spin-$\half$ representation
$\hrho=\shalf(\hat{1}+\vec{\rho}\cdot\vec{\hat{\sigma}})$. 
The determination of the unknown polarization vector
$\vec{\rho}$ relies on the coupling of S with the \as ~A, which we first
take as another two-level system $(n=2)$. The observables
$\hat{\omega}$ and $\hat{o}$ to be measured are the $z$-components
$\siz$ and $\sz$ of S and A, which may be equal to $1$ or $-1$.
The projection operators are $\hat{\pi}_i=\half(\hat{1}\pm \siz)$ and
$\hat{p}_a=\half(\hat{1}\pm \sz)$ for $i$ and $a$ equal to $\pm
1$. Experiments determine the four joint probabilities 
$P_{\alpha}=\{P_{++}, P_{+-}, P_{-+}, P_{--},\}$ for
$\sigma_3$ and $s_3$ to equal $1$ or $-1$.  These probabilities are
related to the three real parameters $\vec{\rho}$ of $\hrho$
through Eq.~(\ref{2}), which reads
\BEA
\label{3}
P_{\alpha}=u_{\alpha}+\vec{v}_{\alpha}\cdot\vec{\rho},~~~~~~~~~~~~~~~~~~~~~~\\
u_\alpha =\shalf\left[
\hU(\hat{1}\otimes \hr)\hU^\dagger
\right]_{\alpha,\alpha},~~
\vec{v}_{\al}=\shalf\left[
\hU(\hat{\vec{\sigma}}\otimes\hat{r})\hU^\dagger 
\right]_{\alpha,\alpha}, 
\label{4}
\EEA
with  $\alpha=\{ia\}=\{\pm\pm\}$ and matrix elements  taken
in the standard representation of the Pauli matrices $\hat{\vec{\sigma}}$
and $\hat{\vec{s}}$. 

By construction, the mapping (\ref{3}, \ref{4}) is such 
the the probabilities $P_{\alpha}$ are non-negative and normalized
for any $\hrho$ such that $\vec{\rho}\,^2\leq 1$. These properties are 
expressed by
\BEA
\label{5-}
u_\alpha\geq |\vec{v}_\alpha|,~~~~~~~~~
\sum_{\alpha}u_{\alpha}=1,\quad 
\sum_{\alpha}\vec{v}_{\alpha}=0.
\label{5}
\EEA
The determinant $\Delta$ of the transformation $\hrho\mapsto P_{\alpha}$ is 
four times the volume of the parallelepiped having any three of the four vectors
$\vec{v}_{\alpha}$ as its sides, e.g.,
$\Delta=4\vec{v}_{++}\cdot(\vec{v}_{-+}\times\vec{v}_{+-})$.  Provided the
evolution operator $\hU$ is such that the vectors $\vec{v}_{\alpha}$ are
not coplanar, the transformation (\ref{3}) can be inverted, and
$\hrho$ is deduced from the set $P_{\alpha}$ of {\it classical
probabilities}.  Alternatively, $\hrho$ is deduced from the knowledge
of the expectation values $\langle \hat{\sigma}_3\rangle$, $\langle
\hat{s}_3\rangle$ and $\langle \hat{\sigma}_3\hat{s}_3\rangle$
at the time $t=\tau$, which are
simultaneously measurable and are in one-to-one correspondence with
the set $P_{\alpha}$.

We first look for the upper bound of $|\Delta|$ implied by
the conditions (\ref{5}). First we note that $|\Delta|$
increases with $|\vec{v}_\alpha|$ for each $\al$. We therefore maximize 
$\Delta^2$ under the constraints $\sum_\al |\vec{v}_\al|=1$
and $\sum_\al \vec{v}_\al=0$, that we account for by means of 
Lagrange multipliers $\lambda$ and $\vec{\mu}$.
Varying ${\tiny \frac{1}{8}}\Delta^2+\lambda\sum_\alpha |\vec{v}_\alpha|
+\vec{\mu}\sum_\alpha\vec{v}_\alpha$ we find
\BEA
\Delta(\vec{v}_{+-}\times \vec{v}_{-+})=\lambda
\left(
\frac{\vec{v}_{++}}{|\vec{v}_{++}|}
-\frac{\vec{v}_{--}}{|\vec{v}_{--}|}
\right)
\EEA
and other equations resulting from all permutations of the $\al$'s.
This yields symmetric solutions for which the four vectors $\vec{v}_\al$,
$\al=\{\pm\pm\}$ form a regular tetrahedron: 
\BEA 
\label{tetra}
u_{\al}=|\vec{v}_{\al}|=\frac{1}{4},\quad
\frac{\vec{v}_{\al}\cdot\vec{v}_{\beta}} 
{|\vec{v}_{\al}|\,|\vec{v}_{\beta}|}
=-\frac{1}{3},\quad \al\not=\beta.  
\EEA 
As could have been anticipated, the solutions are not unique: they
follow one from another by rotation in the space of the spins
$\vec{\sigma}$ and permutations of the indices $\al$. The
corresponding upper bound for the modulus of the determinant is 
$|\Delta|=1/(12\sqrt{3})$.

Let us show that this upper bound can be reached, provided the initial
state of the assistant is {\it pure}.  We have to construct a unitary
operator $\hat{U}$ that satisfies Eqs.~(\ref{4}, \ref{tetra}), and to
find a Hamiltonian $\hat{H}$ and an interaction time $\tau$ such that
$\hU=e^{-i\hH\tau}$.  Let us exhibit an example of such a solution.
We assume that the assistant is initially polarized in the $z$-direction,
$\hat{r}=\hat{p}_+=\shalf(1+\hat\sigma_3)$, and we orient 
the tetrahedron $\vec{v}_\al$ in the direction 
$\vec{v}_{+\pm}=(\pm 1,1,\pm 1)/4\sqrt{3}$, 
$\vec{v}_{-\pm}= (\pm 1,-1,\mp1)/4\sqrt{3}$.
The correspondence (\ref{3}) then takes an especially simple form:
\BEA
\label{6}
\rho_1=\sqrt{3}\langle\hat{s}_3\rangle,\quad
\rho_2=\sqrt{3}\langle\hat{\sigma}_3\rangle,\quad
\rho_3=\sqrt{3}\langle\hat{\sigma}_3\,\hat{s}_3\rangle,
\EEA
yielding directly the density matrix $\hrho$ in terms of the 
expectation values and the correlation of the commuting observables 
$\hat{\sigma}_3$ and $\hat{s}_3$ in the final state.
It is easy to verify that this correspondence can be achieved
under the action of the Hamiltonian
\BEA
\label{unop} 
\hat{H}=\six\frac{\sx\cos\phi+\sz\sin\phi}{\sqrt{2}}
+\frac{(\sy-\sx)\sin\phi+\sz\cos\phi}{2},
\label{7}
\EEA
where $2\phi=0.95531$ is the angle between $\vec{v}_{++}$ and the $z$-axis,
that is, $\cos 2\phi=1/\sqrt{3}$. Noting that 
$\hat{H}^2=\sin^2\chi$, where $\chi=1.11069$ satisfies 
$\cos\chi=\half\cos\phi$,  we obtain
$\hU$$=$${\rm e}^{-{\rm i}\hH\tau}
=\cos(\tau\sin\chi)-i\hat{H}\sin(\tau\sin\chi)/\sin\chi$. 
Taking as duration of the evolution $\tau=\chi/\sin\chi$, we obtain
$\hU=\cos\chi-i\hat{H}$. Insertion in (\ref{4}) allows to check Eq.~(\ref{tetra})
and to get the expected optimal correspondence (\ref{6}).

The simpler Hamiltonian
\BEA\label{artashir}
\hat{H}={\tiny\frac{1}{\sqrt{2}}}\,
\six\sx+{\tiny \shalf}(\sy\sin\phi+\sz),
\EEA
which results from (\ref{7}) through rotation of $\hat{\vec{s}}$,
also achieves an optimal mapping $\hrho\mapsto P_\al$, provided
$\hat s_3$ is replaced by $\hat s_1\sin\phi+\hat s_3\cos\phi$ both in the 
measured projections $\hat p_a=\shalf(1\pm\hat s_3)$ and in the initial state
$\hat r=\hat p_+$.
The first term of Eq. (\ref{artashir}) describes, in the spin language, 
an {\it Ising coupling}, while the second term represents a
{\it transverse magnetic field} acting on A only.

{\it Larger assistant.} We have optimized above the determination of
$\hrho$ by coupling S to an assistant A that starts in a pure state and has
the same dimension $n=m=2$ as S. It is natural to wonder whether the
quality of the measurement, as expressed by the magnitude of $\Delta$,
may be improved for $n>m=2$ and/or for mixed initial states of A.  We
may, for instance, consider an \as~ consisting of $q$ spins, in which
case $m=2$ and $n=2^q$. We now denote as $\hat{s}_3$ some two-valued
observable of A which is subjected to measurement at the time
$\tau$. The only changes in (\ref{2}) are the dimension $n$ of the
matrix $\hr$ and the fact that the two projection operators
$\hat{p}_a$ no longer constitute a complete set in the Hilbert space
of A. Experiment still provides the four probabilities $P_{\al}=\tr
\,\hR_\tau (\hat{\pi}_i \otimes\hat{p}_a)$ with unit sum, where
$\hR_\tau=\hU(\hrho\otimes\hr)\hU^\dagger$ is the final density
operator in the $mn$-dimensional space of S+A, and the mapping
$\hrho\mapsto P_\al$ keeps the form (\ref{3}).  The conditions 
(\ref{5}) still hold, since they express simply that the correspondence
(\ref{2}) or (\ref{3}) preserves the positivity and the
normalization. When obtaining the upper bound $1/(12\sqrt{3})$ for
$|\Delta|$ we relied only on these conditions. Therefore, using a larger 
\as~ {\it cannot improve} upon the optimal solutions found for $n=2$ and
pure $\hr$.  

In all cases, the maximum of $|\Delta|$ is reached for
mappings (\ref{3}, \ref{4}) which involve the regular tetrahedron
(\ref{tetra}).  In such mappings there exist 4 pure states,
$\hrho_{\bar\alpha}=\shalf(1-4\vec{v}_{\bar\al}\cdot \hat{\vec{\sigma}})$, 
for which one probability, $P_{\bar\al}$, vanishes.

{\it Completely disordered assistant.} Returning to the case $n=m=2$
we saw that we could reach the upper bound $1/(12\sqrt{3})$ of
$|\Delta|$ if the assistant is initially in a pure state. In order to
explore how much is lost if it is in a mixed state, we consider the
extreme situation in which $\hr=\half\hat{1}$ is the completely
disordered state. There is an advantage in using such a state, as it
is easier to prepare than a pure state: one lets the \as~ interact
with a hot thermal bath; for a spin, one leaves it
unpolarized. Eqs.~(\ref{3}, \ref{4}) taken for $\hr=\half\hat{1}$ show
that too simple evolutions may lead to a vanishing determinant
$\Delta$.  For instance, the above evolution (\ref{7}), which we
optimized for $\hat{r}=\hat{p}_+$, is completely ineffective for
$\hr=\half\hat{1}$, since it maps any $\hrho$ onto the trivial set of
probabilities $P_\al=\frac{1}{4}$.

In order to maximize $|\Delta| $ with respect to $\hat U$ for
$\hr=\half\hat{1}$, we now have to evaluate the vectors $\vec{v}_\al$
from (\ref{4}). We only sketch this calculation here. We represent
$U_{ai,bj}$ as $U_{+\cdot,+\cdot}=VK$, $U_{-\,\cdot,+\,\cdot}=WK'$,
$U_{+\cdot,-\cdot}=VK'X$, $U_{-\cdot,-\cdot}=-WKX$,
in terms of two hermitian positive  $2\times 2$ matrices $K$ and
$K'$ such that $K^2+K'^2=1$, and of three unitary $2\times 2$ matrices
$V$, $W$, $X$ in the space $i=\pm$, $j=\pm$ of S. Using the invariances of the
problem, we may without restrictions parametrize $K=K(\theta,\varphi)\equiv
\cos\theta\cos\varphi+\sin\theta\sin\varphi\,\vec{\chi}
\cdot\vec{\hat{\sigma}}$ with unit vector $\vec\chi$, implying
$K'=K(\theta-\frac{\pi}{2},\varphi)$ and $K^2-{K'}\,^2=K(2\theta,2\varphi)$.
We characterize $V$ by 
$V^\dagger\hat{\sigma}_3V=\vec{\eta}\cdot\vec{\hat{\sigma}}$,
$W$ by $W^\dagger\hat{\sigma}_3W=\vec{\zeta}\cdot\vec{\hat{\sigma}}$,
and take $X=\vec{\xi}\cdot\vec{\hat{\sigma}}$, where also
$\vec{\eta}$, $\vec{\zeta}$, $\vec{\xi}$ are unit vectors, with
$\vec{\chi}\cdot\vec{\xi}=0$. We then obtain $u_\al=\frac{1}{4}$ and
the $\vec{v}_\al$, whence
\BEA
\Delta=\frac{1}{32}
\sin 4\theta\sin 4\varphi\,(\vec{\chi}\times\vec{\xi})\cdot
\left[\,
(\vec{\zeta}\cdot\vec{\xi})\,\vec{\eta}+
(\vec{\eta}\cdot\vec{\xi})\,\vec{\zeta}\,
\right].~
\label{111}
\EEA
We now determine the maximum value of $|\Delta|$
attainable for $\hr=\half\hat{1}$ and the corresponding optimal evolution
$\hat{U}$. The best angles are $\theta=\varphi=\pi/8$.
Choosing $\vec\chi=(1,0,0)$ and $\vec\xi=(0,1,0)$, we  see that the
largest $|\Delta|$ is $1/32$.
This still corresponds to 
$\vec{v}_\al$ forming a regular tetrahedron, but with $|\vec{v}_\al|=
\sqrt{3}/8$ instead of $|\vec{v}_\al|=\frac{1}{4}$. Accordingly, 
for the same orientation of the ${\vec{v}}_\al$'s, the factor
$\sqrt{3}$ in (\ref{6}) is replaced by $2$. Rather surprisingly, the
efficiency of the new scheme is not much worse than when the \as~ is
prepared in a pure state. However, the Hamiltonians needed now to maximize
$\Delta$ require a more complicated coupling than in (\ref{7}). 
Among various possibilities, we present here an example:
\BEA
\label{8}
\hH=\half\vec{\hat{\sigma}}\cdot\vec{\hat{s}}
-\sqrt{2}\siy\sy
+\frac{1}{\sqrt{2}}(\six+\sx)
\EEA
(the signs of the three terms can be changed independently).
This Hamiltonian involves an {\it anisotropic Heisenberg} interaction
and an {\it external field} acting {\it symmetrically} on S and A.
Using Eq.~(\ref{4}) with $\hr=\half\hat{1}$, one can check that 
for $\tau=\frac{1}{4}(2k+1)\pi$ the evolution operator
$\hU={\rm e}^{-i\hH\tau}$ leads to an optimal solution with $\Delta=
1/32$. For $\tau=\pi/4$ Eq.~(\ref{6}) is replaced by
\BEA
\label{9}
\rho_1=2\langle\hat{\sigma}_3\hat{s}_3\rangle,\quad
\rho_2=2(\langle\hat{s}_3\rangle\cos\gamma
+\langle\hat{\sigma}_3\rangle\sin\gamma),\\
\rho_3=2(\langle\hat{\sigma}_3\rangle\cos\gamma
-\langle\hat{s}_3\rangle\sin\gamma),\quad
\gamma=\frac{\pi(1+\sqrt{2})}{4}.
\EEA

{\it Conclusion.} The {\it non-commutative information} contained in
the density matrix of quantum system can be transformed by a
one-to-one correspondence into ordinary information associated with a
set $P_\al$ of {\it ordinary probabilities for exclusive events}. 
The price to be paid is
the introduction of our extra \as~ system A. This correspondence can
be experimentally implemented by  repeatedly letting S and A suitably interact,
then by performing each time simultaneous measurements of two
commuting observables $\ho$ and $\hat{o}$ pertaining to S and A,
respectively. Counting of events yields the probabilities $P_{ia}$ 
for $\ho$ to take the value $\omega_i$ and for $\hat{o}$
to take the value $o_a$. Provided $\ho$ is non degenerate so that its
number of distinct eigenvalues $\omega_i$ ($1\leq i\leq m$) equals the
dimension $m$ of the Hilbert space of S, and provided the number of
eigenvalues of $\hat{o}$ is also $m$ (at least), the correspondence
$\hrho\mapsto P_\al$ can be inverted, as we displayed on several
examples. The second condition implies that the \as~ has a dimension
$n\geq m$.  Hence an {\it initially unknown} $\hrho$ can be determined
via $P_\al$ by means of a {\it single apparatus}. 

As compared to the
standard determination schemes of $\hrho$ based on direct
non-commutative measurements of S, the present method has several
advantages. {\it i)} It is more economical, since it involves only one
observable $\ho$ of S and one observable $\hat{o}$ of A, whereas direct
determinations require measuring at least $m+1$ non-commuting
observables of S. {\it ii)} This full set of $m+1$ observables is not
always accessible in practice. For instance, for a two-level atom
prepared in some unknown state $\hat\rho$, $\rho_3$ is readily measured
through the occupation probability of the excited state, but $\rho_1$
and $\rho_2$ can be determined only indirectly. Interaction of S with
another initially known two-level atom A (whose own preparation may be
straightforward, as we saw above) may provide the full $\hrho$
through mere simultaneous measurements of the occupation probabilities
for the levels of S and A. {\it iii)} It has been stressed \cite{swedish} that the
use of standard statistical and information theoretical methods for
dealing with incomplete or noisy experimental data cannot be directly
extended to quantum mechanics, because results are produced there by means
of non-commutative measurements; indeed, these data pertain to
different contexts as they are produced by different apparatuses. The
present scheme, involving only commutative measurements,
circumvents this difficulty.

Taking as a criterion of quality of our measurement schemes the size
of the determinant of the mapping $\hrho\mapsto P_\al$, we have explored
for $m=2$ the conditions that lead to the {\it best determination} of
$\hrho$ for some uncertainty on the set $P_\al$.  For an \as~ with
dimension $2$, its known initial state $\hr$ should be pure and the
parameters of the Hamiltonian should be suitably chosen.  An example
of an optimal evolution is generated by the Hamiltonian (\ref{7})
or (\ref{artashir}).  We
have also seen that the determinant cannot be enlarged by use of an
\as~ with dimension larger than $2$. (However, $\hat{o}$ may then take
more than two eigenvalues $o_a$ and the probabilities $P_{ia}$ become
more numerous than needed; the fact that they are related to one
another independently of $\hrho$ opens the possibility of improving
the determination of $\hrho$ through cross-check of the data $P_\al$.)

For $m=2$, the optimal mappings (\ref{3}), (\ref{tetra}) amount to
identify, via a dynamical process, the joint probabilities $P_\al$ for 
$\hat\sigma_3$ and $\hat s_3$ with the expectation values in the state 
$\rho$ of the observables $\half(\hat 1-\hat\Omega_\al)$, where
$\hat\Omega_\al=\half(\hat 1-4\vec{v}_\al\cdot \vec{\hat{\sigma}})$
pertains to the system S. The 4 observables $\hat\Omega_\al$
are projection operators, satisfying ${\rm tr}\,\hat\Omega_\al\hat\Omega_\beta=
\frac{1}{3}$ for $\al\neq\beta$, and spanning the space of observables 
$\hat\omega$ of S. 
For $m>2$ we conjecture that a bound on $\Delta$ may be found by considering
in the Hilbert space of S a set of $m^2$ projections $\hat\Omega_\al$,
satisfying ${\rm tr}\,\hat\Omega_\al=1$, ${\rm tr}\,\hat\Omega_\al\hat\Omega_\beta=
1/(m+1)$ for $\al\neq\beta$, $\sum_\al \hat\Omega_\al=m\,\hat 1$, and
constituting a basis for the observables $\hat\omega$. Then the mapping
matrix $R$ in  $P_\alpha= \sum_{ij}R_{\al,ij}\rho_{ji}$ is expected
to be given by $m(m-1)R_{\al,ij}=\delta_{ij}-\Omega_{\al,ij}$.
This form makes the determinant $\Delta$ stationary under the constraints 
imposed by positivity and normalization alone. As above, there
are $m^2$ pure states $\hat\rho_{\bar\al}=\hat\Omega_{\bar\al}$ for which
one probability, $P_{\bar\al}$, vanishes. This conjecture yields for 
$\Delta^2$ the upper bound $m^{-1}[m(m+1)(m-1)^2]^{1-m^2}$,
which generalizes the $m=2$-result $\Delta^2=1/(12\sqrt{3})^2$.

It might have been expected that a completely disordered \as~ makes
the determination of $\hrho$ from observation of $P_\alpha$ unprecise or
even infeasible; for $m=2$ it turns out that the best Jacobian in this 
case is smaller than the maximum one only by a factor $(\sqrt{3}/2)^3
\simeq 0.65$, and that Hamiltonians as simple as (\ref{8}) can be used.

Indeed, the various types of two-level systems on which
experiments are currently performed (NMR, quantum and atomic optics,
spintronics) feature Hamiltonians similar to 
 (\ref{artashir}) and (\ref{8}) that optimize the
process, with Ising or Heisenberg types of couplings. For instance,
the spin-spin interaction between two single-electron quantum dots is
usually anisotropic due to spin-orbit coupling or to a lack of
symmetry of the host material; see \cite{ka} for a recent
discussion. Experiments can therefore easily be designed along the
above ideas. They will demonstrate that the principle of
complementarity, which seems to imply that different measurement
devices are needed to fully determine a quantum state, can be
by-passed by using an \as, even completely disordered.

The work of A.E. A is part of the research programme of the Stichting voor
Fundamenteel Onderzoek der Materie (FOM, financially supported by
the Nederlandse Organisatie voor Wetenschappelijk Onderzoek (NWO)).
R. B. acknowledges hospitality at the University of Amsterdam,
and A.E. A. and Th.M. N. at the CEA-Saclay.

\end{document}